\documentclass[aps,pra,showpacs]{revtex4}

\usepackage{graphicx}
\usepackage{amsmath}
\usepackage{amssymb}
\usepackage{amsthm}

\begin{document}
\title{One-dimensional Coulomb problem in Dirac materials}

\author{C. A. Downing}
\affiliation{School of Physics, University of Exeter, Stocker Road,
Exeter EX4 4QL, United Kingdom}

\author{M. E. Portnoi}
\email[]{m.e.portnoi@exeter.ac.uk}
\affiliation{School of Physics,
University of Exeter, Stocker Road, Exeter EX4 4QL, United Kingdom}
\affiliation{International Institute of Physics, Universidade Federal do Rio Grande do Norte, Natal - RN, Brazil}

\date{November 16, 2014}

\begin{abstract}
We investigate the one-dimensional Coulomb potential with application to a class of quasi-relativistic systems, so-called Dirac-Weyl materials, described by matrix Hamiltonians. We obtain the exact solution of the shifted and truncated Coulomb problems, with the wavefunctions expressed in terms of special functions (namely Whittaker functions), whilst the energy spectrum must be determined via solutions to transcendental equations. Most notably, there are critical bandgaps below which certain low-lying quantum states are missing in a manifestation of atomic collapse.
\end{abstract}

\maketitle

\section{\label{intro}Introduction}

The Coulomb problem in quantum theory is a historic problem of theoretical physics \cite{Bohr}. Its solution, which can be written down analytically, is a cornerstone of quantum mechanics and gives tremendous insight into the hydrogen atom \cite{Schrodinger, Pauli, Landau1}. Moreover, its solution in reduced dimensions is also highly significant: experiments with electrons confined to a plane led to considerations of the Coulomb problem in two dimensions (2D) \cite{Flugge, Yang, Parfitt, Makowski}, whilst the history of the one-dimensional (1D) Coulomb problem is long, interesting and sometimes controversial \cite{Loudon, Andrews, Haines}.

The analogous relativistic problem \cite{Dirac, Gordon, Landau2} as governed by Dirac's equation, is equally fascinating and likewise the problem has also been investigated in low dimensions, both in 2D \cite{Guo, Dong} and 1D \cite{Krainov, Spector, Entin}. The rise of Dirac materials \cite{Wehling}, condensed matter systems with quasi-particles well-described by the Dirac equation, has led to revisits of Dirac-Kepler problems with Dirac-like matrix Hamiltonians. One example is the 2D relativistic solution and its application to graphene \cite{Shytov}, which has charge carriers described by a massless Dirac-Weyl equation. Graphene, a single atomic layer of carbon atoms in a honeycomb lattice \cite{CastroNeto}, is the star of the Dirac materials; however, there are in fact a plethora of other materials such as topological insulators \cite{Hasan, Qi}, transition metal dichalcogenides \cite{Xiao}, carbon nanotubes \cite{Charlier} and 3D Weyl semimetals \cite{Young}, which provide physicists  a new playground to investigate quasi-relativistic phenomena.

Here we look at the quasi-relativistic Coulomb problem in 1D at the level of a two-by-two Dirac-like matrix Hamiltonian. Our results should be useful in several areas for various quasi-1D Dirac systems, most notably narrow-gap carbon nanotubes and graphene nanoribbons, for example: in the understanding of the energy spectra of donors and excitons, table-top experiments on atomic collapse, vacuum polarization effects, Sommerfeld factor and the suppression of van Hove singularities, Coulomb blockade and zero-bias anomalies, magnetoexcitons and so on. Besides, the intrinsic beauty of analytic results in quantum mechanics is almost always coupled with greater insight, as well as being sturdy platforms on which to test new numerical methods or perturbative schemes.

The low-energy spectrum of a typical 1D Dirac material can be described by a single-particle matrix Hamiltonian
\begin{equation}
\label{ham1}
\hat{H}_1 = v_F \begin{pmatrix} 0 &  \hat{p}_{x}-i\hbar \Delta \\ \hat{p}_{x}+i\hbar \Delta & 0 \end{pmatrix} + U(x)
\end{equation}
where $v_F$ is the Fermi velocity (which can be for example $v_F \approx c/300$ for carbon nanotubes or graphene nanoribbons), $2 \hbar v_F |\Delta|$ is the bandgap and the momentum operator $\hat{p}_{x}$ acts along the axis of the effectively 1D system. The same Hamiltonian, Eq.~\eqref{ham1}, describes a 2D Weyl material, e.g. graphene or the surface of a topological insulator, subjected to a 1D potential $V(x)$ constant in the $y$-direction, in which case $\Delta \to k_y$ \cite{Yung}. We make the unitary transform $U = \tfrac{1}{\sqrt{2}} \bigl(\begin{smallmatrix} 1&1\\ 1&-1 \end{smallmatrix} \bigr)$ with Eq.~\eqref{ham1} and obtain the following system of equations
\begin{equation}
\label{coulomb1}
  \begin{pmatrix} \partial_{x} & -\Delta \\ \Delta & -\partial_{x} \end{pmatrix}
   \left(
 \begin{array}{c}
  \psi_{1}(x)  \\
	 \psi_{2}(x)
 \end{array}
\right)
	= i(\varepsilon - V(x))  \left( \begin{array}{c}
  \psi_{1}(x)  \\
	 \psi_{2}(x)
 \end{array} \right).
\end{equation}
where we have scaled the eigenvalue $\varepsilon = E/ \hbar v_F$ and potential energy $V(x)=U(x)/ \hbar v_F$.

In what follows we investigate the quasi-1D Coulomb potential with two different modifications, the so-called `shifted' and `truncated' Coulomb problems. Both modifications introduce a regularization scheme at the origin, so as to avoid problematic boundary conditions well known in the non-relativistic case \cite{Loudon} and more importantly to be more physically meaningful. A cut-off naturally arises in nanotubes and quantum wires due to the finite (albeit small) size of the quantum confined direction, which is related to the radius of the wire \cite{Banyai}. The third main alteration to the Coulomb potential is the Ohno potential \cite{Perebeinos, Note}, but we omit a treatment of this case as it is only quasi-exactly solvable \cite{Turbiner} in terms of confluent Heun functions \cite{Downing}.

For completeness, we note that exponentially decaying potentials, which are of a short-range nature, have also been considered in quasi-1D Dirac systems in various forms \cite{Adame, Stone, Hartmann}. However, it is the pure Coulombic long-range interaction, decreasing like the inverse of separation, which is the subject of this work as it is well known that screening is suppressed in low-dimensional systems \cite{Haug}. Indeed, in the case of 2D Dirac-Weyl systems like graphene screening does not alter the long range functional dependence of the Coulomb interaction \cite{CastroNeto}, whilst screening is further reduced in carbon nanotubes \cite{Charlier}. Additionally, in a similar framework to this work, transmission problems through periodic potentials \cite{Barbier} as well as linear \cite{Cheianov} and smooth step potentials \cite{Reijnders} have been treated.

Furthermore, it should be mentioned that the confinement of Dirac-like particles in one-dimensional potentials is the subject of considerable recent attention from the applied mathematics community \cite{Zhong, Downes, Pankrashkin, Jakubsky}. Our results here, using two explicit toy models of non-integrable potentials, provide a complementary approach both more accessible to physicists and closer to experimental reality.

\section{\label{sec2}The shifted Coulomb problem}

\begin{figure}[htbp]
 \includegraphics[width=0.3\textwidth]{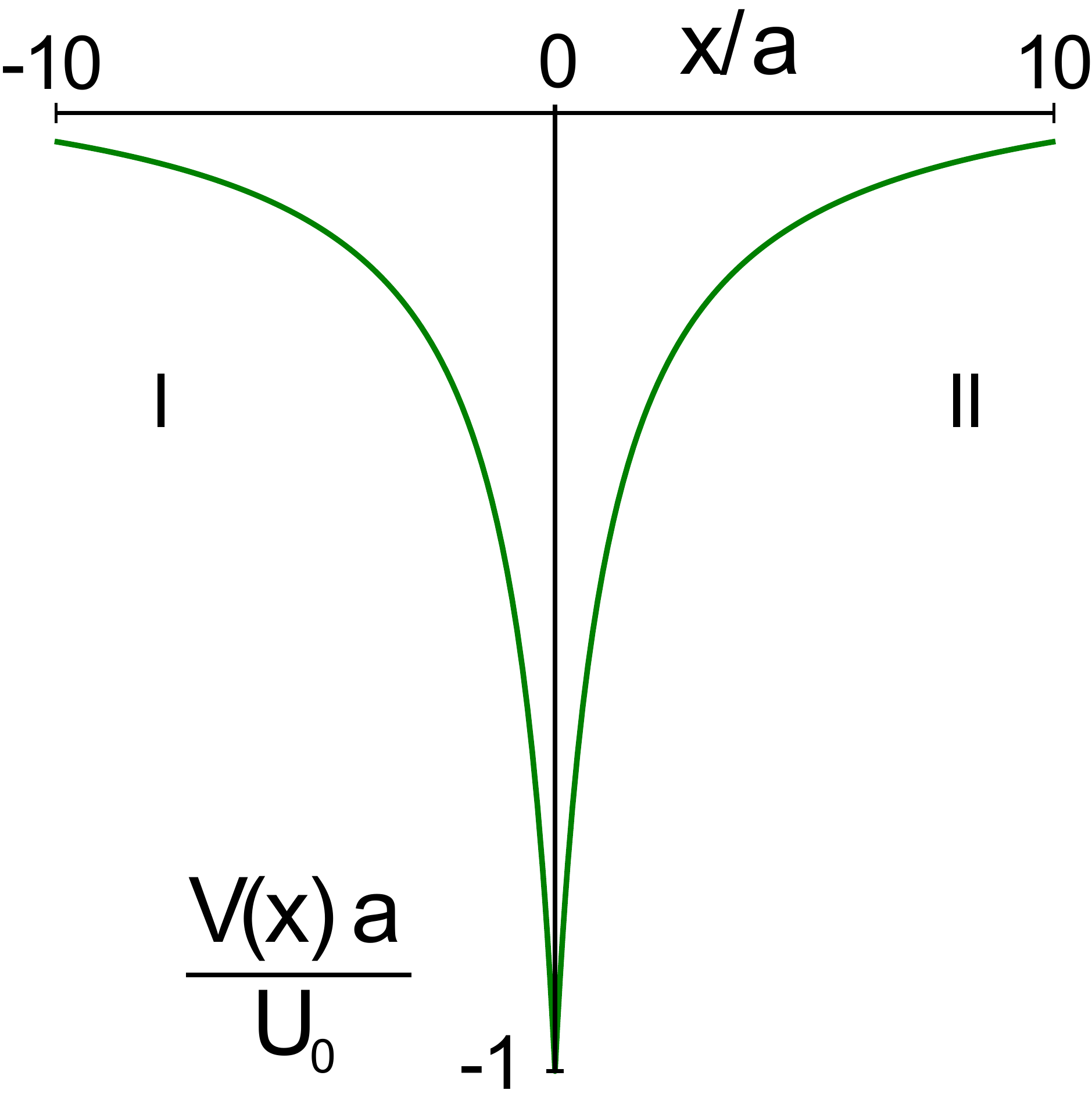}
 \caption{A plot of the shifted Coulomb potential, defined by Eq.~\eqref{pot1}.}
 \label{shift3}
\end{figure}

In this section we shall investigate the shifted 1D Coulomb potential plotted in Fig.~\ref{shift3} and explicitly given by
\begin{equation}
\label{pot1}
	V_{s}(x) = \frac{-U_0}{a+|x|},
\end{equation}
where $a$ is the shift length, and the dimensionless number $U_0 = \tfrac{e^2}{4 \pi \epsilon} \tfrac{1}{\hbar c} \tfrac{c}{v_F}$ is an effective fine structure constant, which in the case of carbon nanotubes or graphene nanoribbons is $U_0 \approx \tfrac{300}{137}$.

Upon substitution of Eq.~\eqref{pot1} into Eq.~\eqref{coulomb1}, the wavefunction component $\psi_{1}(x)$ in the region II $(x>0)$ satisfies a modified form of the confluent hypergeometric equation, called the Whittaker differential equation, in the variable $\xi = 2 \kappa (a+x)$,
\begin{equation}
\label{coulomb3}
 \frac{d^2}{d\xi^2}\psi_{1}(\xi) + \left( -\frac{1}{4} + \frac{\mu}{\xi} + \frac{1/4-\nu^2}{\xi^2}  \right) \psi_{1}(\xi) = 0,
\end{equation}
where
\begin{equation}
\label{coulomb4}
 \mu = \frac{\varepsilon U_0}{\kappa}, \quad \nu = i U_0 - \frac{1}{2},
\end{equation}
with $\kappa = \sqrt{\Delta^2-\varepsilon^2}>0$, as we consider bound states ($ |\varepsilon| < |\Delta|$) only. An asymptotically convergent solution can be constructed, known as the Whittaker function of the second kind \cite{Gradshteyn}
\begin{equation}
\label{coulomb40}
 W_{\mu, \nu} (\xi) = \xi^{1/2+\nu} e^{-\xi/2} U\left( \tfrac{1}{2} + \nu - \mu, 1 + 2 \nu, \xi \right),
\end{equation}
where the Tricomi function $U(\alpha, \beta, \xi)$ is built from a linear combination of the usual confluent hypergeometric functions of the first kind:
\begin{eqnarray}
\label{coulomb41}
 U(\alpha, \beta, \xi) = \frac{\Gamma{(1-\beta)}}{\Gamma(\alpha-\beta+1)} F(\alpha, \beta, \xi)
 \nonumber \\
		+ \frac{\Gamma{(\beta-1)}}{\Gamma(\alpha)} \xi^{1-\beta} F(\alpha-\beta+1, 2-\beta, \xi),
\end{eqnarray}
where $F(\alpha, \beta, \xi)$ is a hypergeometric series given by
\begin{equation}
\label{coulomb42}
 F(\alpha, \beta, \xi) = 1 + \frac{\alpha}{\beta} \xi + \frac{\alpha (\alpha + 1)}{\beta (\beta + 1)} \frac{\xi^2}{2!} + ...
\end{equation}
This construction ensures the desired decaying behavior at infinity: $U(\alpha, \beta, \xi) \to \xi^{-\alpha}$.

\begin{figure}[htbp]
 \includegraphics[width=0.4\textwidth]{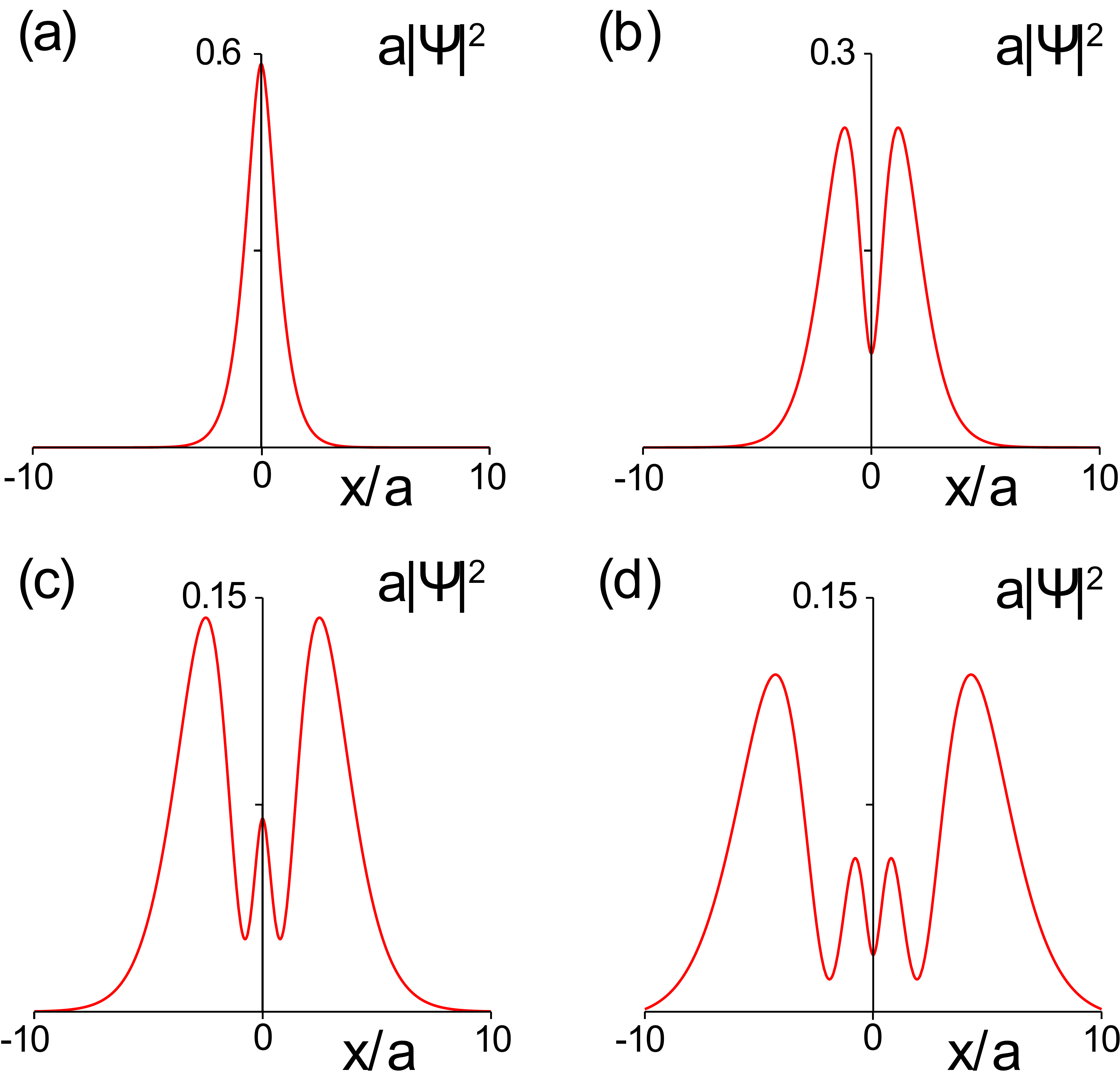}
 \caption{Probability density plots of the first four successive bound states with bandgap $\Delta a = 1$ and potential strength $U_0 = \tfrac{300}{137}$, where $(a)~\varepsilon a = -0.248$, $(b)~\varepsilon a = 0.350$, $(c)~\varepsilon a = 0.570$ and $(d)~\varepsilon a = 0.703$.}
 \label{loudon1}
\end{figure}

One can then proceed to find the full solution to the system of equations~\eqref{coulomb1}: in region II ($x>0$) we obtain
\begin{equation}
\label{coulomb5}
\Psi_{II}(x) =  \frac{c_{II}}{\sqrt{a}} \left(
 \begin{array}{c}
  W_{\mu, \nu} (\xi_{II}) \\
	 -\frac{\kappa+i \varepsilon}{\Delta} W_{\mu, \nu+1} (\xi_{II})
 \end{array}
\right),
\end{equation}
similarly in region I ($x<0$) it follows
\begin{equation}
\label{coulomb6}
\Psi_{I}(x) =  \frac{c_{I}}{\sqrt{a}} \left(
 \begin{array}{c}
   \frac{\kappa +i \varepsilon}{\Delta} W_{\mu, \nu+1} (\xi_{I}) \\
	 W_{\mu, \nu} (\xi_{I})
 \end{array}
\right),
\end{equation}
where now the variable $\xi_{I, II} = 2 \kappa (a \mp x)$.

Using the continuity condition for both wavefunction components $ \psi^{I}_{1,2}|_{x=0-} = \psi^{II}_{1,2}|_{x=0+}$ with Eq.~\eqref{coulomb5} and Eq.~\eqref{coulomb6}, yields the ratio of constants $c_{II}/c_I = \pm i$, where $c_I$ is found via the normalization condition for a spinor wavefunction
\begin{equation}
\label{coulomb71}
  \int_{-\infty}^{\infty} \left( |\psi_1|^2 + |\psi_2|^2 \right)~\mathrm{d}x = 1.
\end{equation}
Bound state eigenvalues must be determined from the transcendental equation
\begin{equation}
\label{coulomb7}
  \frac{\Delta}{\kappa + i \varepsilon} \frac{W_{\mu, \nu} (2 \kappa a)}{W_{\mu, \nu+1} (2 \kappa a)} = \pm i,
\end{equation}
which can be solved graphically or via other standard root-finding methods. We show in Fig.~\ref{loudon1} four illustrative electron density plots of the lowest bound states for $U_0 = \tfrac{300}{137}$, corresponding to a single charge Coulomb impurity on the axis of a single-walled carbon nanotube, and $\Delta a = 1$. Characteristically, the ground state density has a single peak, followed by two peaks for the first excited state, and so on. The value of the density at the origin alternates from being one of a local maxima to a local minima, but in a noticeable contrast to the non-relativistic case is never zero. This arises from the matrix nature of the Hamiltonian Eq.~\eqref{coulomb1}, which ensures both wavefunction components never vanish simultaneously. Higher energy bound states are more spread in space, with the highest peaks of probability density concentrated in the two outermost shoulders.

\section{\label{sec3}The truncated Coulomb problem}

\begin{figure}[htbp]
 \includegraphics[width=0.3\textwidth]{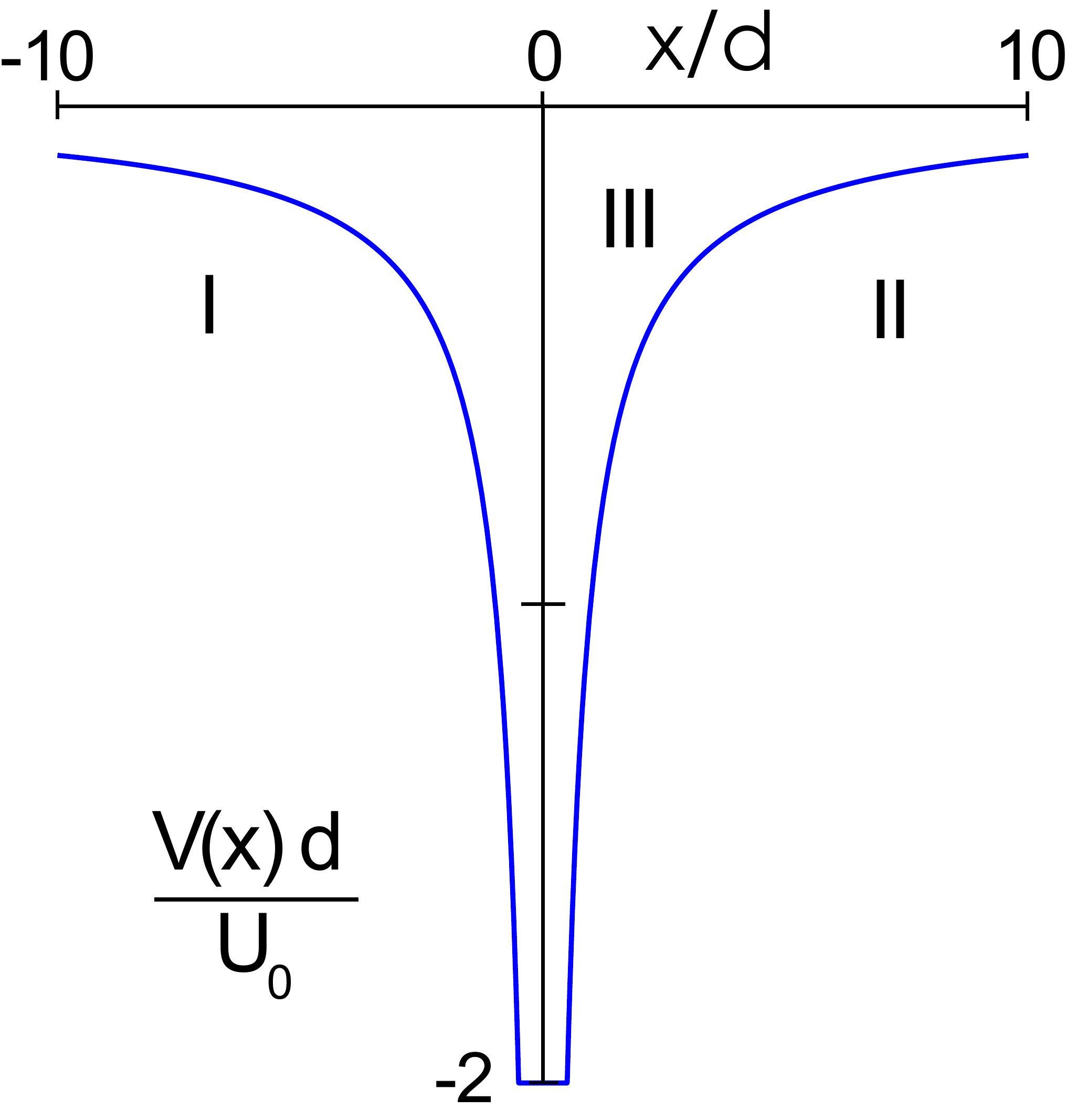}
 \caption{A plot of the truncated Coulomb potential, defined by Eq.~\eqref{pot2}.}
 \label{trunc3}
\end{figure}

We also consider the truncated 1D Coulomb potential, plotted in Fig.~\ref{trunc3} and shaped by the piecewise function
\begin{equation}
\label{pot2}
	V_{t}(x) = \begin{cases} -2 U_0/d, & \mbox{if } |x| \le d/2 \\
						     -U_0/|x|, &\mbox{if } |x| > d/2 \end{cases}
\end{equation}
where the Coulomb potential has been terminated at a radius $d/2$ to form a flat-bottom quantum well at small distances.

In the exterior regions I and II, where $|x|>d/2$, the solutions follow from those of Sec.~\ref{sec2} upon setting $a=0$. In the interior region III, where $|x| \le d/2$, the solutions are simply
\begin{equation}
\label{truncated00}
\Psi_{III}(x) =  \frac{c_{III}}{\sqrt{d}} \left(
 \begin{array}{c}
  \sin (kx) \\
	 f_1(x)
 \end{array}
\right)
 + \frac{c_{IV}}{\sqrt{d}} \left(
 \begin{array}{c}
 \cos (kx) \\
	f_2(x)
 \end{array}
\right),
\end{equation}
where we have introduced the auxiliary two-component function
\begin{equation}
\label{truncated01}
  \left(
 \begin{array}{c}
  f_1(x)  \\
	 f_2(x)
 \end{array}
\right) = \frac{k}{\Delta} \left(
 \begin{array}{c}
  \cos (kx)  \\
	- \sin (kx)
 \end{array}
\right) + \frac{\varepsilon + 2 U_0/d}{i \Delta} \left(
 \begin{array}{c}
  \sin (kx)  \\
  \cos (kx)
 \end{array}
\right),
\end{equation}
which necessitates the introduction of a new wavenumber $k = \sqrt{(\varepsilon+2U_0/d)^2-\Delta^2}>0$, arising from the short-range behavior of the potential. The wavenumber defining the long-range decay of the wavefunction remains $\kappa$, introduced after Eq.~\eqref{coulomb4}. Together, requiring $k, \kappa > 0$, one finds a definite region in which confined states may form, restricted maximally by $|\varepsilon d| < \varepsilon_{max} d = \Delta d$ and minimally by $\varepsilon d> \varepsilon_{min} d = \Delta d - 2U_0$.

\begin{figure}[htbp]
 \includegraphics[width=0.4\textwidth]{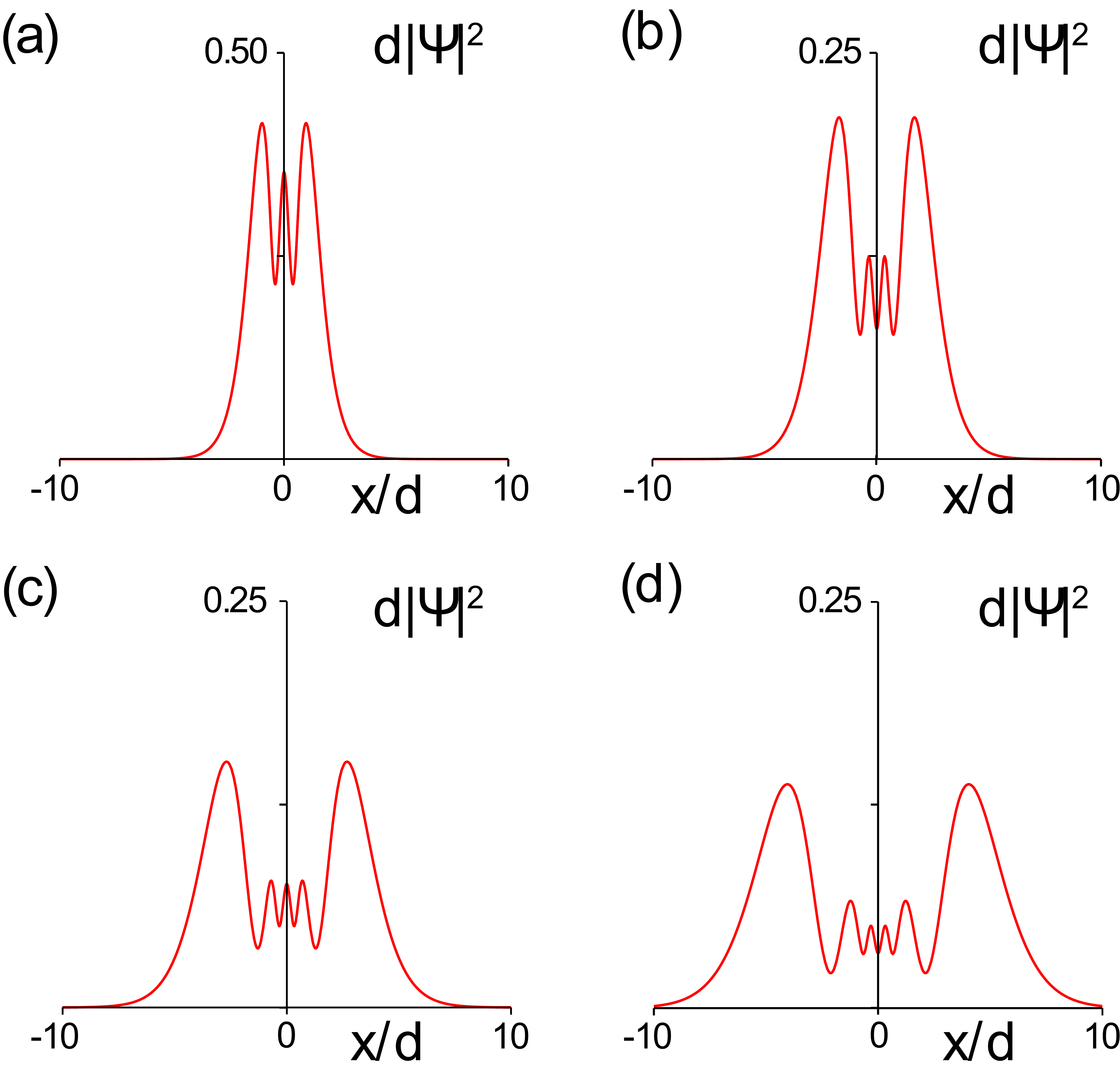}
 \caption{Probability density plots of the first four lowest bound states with bandgap $\Delta d = 1$ and potential strength $U_0 = \tfrac{300}{137}$, where $(a)~\varepsilon d = -0.270$, $(b)~\varepsilon d = 0.191$, $(c)~\varepsilon d = 0.460$ and $(d)~\varepsilon d = 0.623$.}
 \label{stone1}
\end{figure}

\begin{figure}[htbp]
 \includegraphics[width=0.4\textwidth]{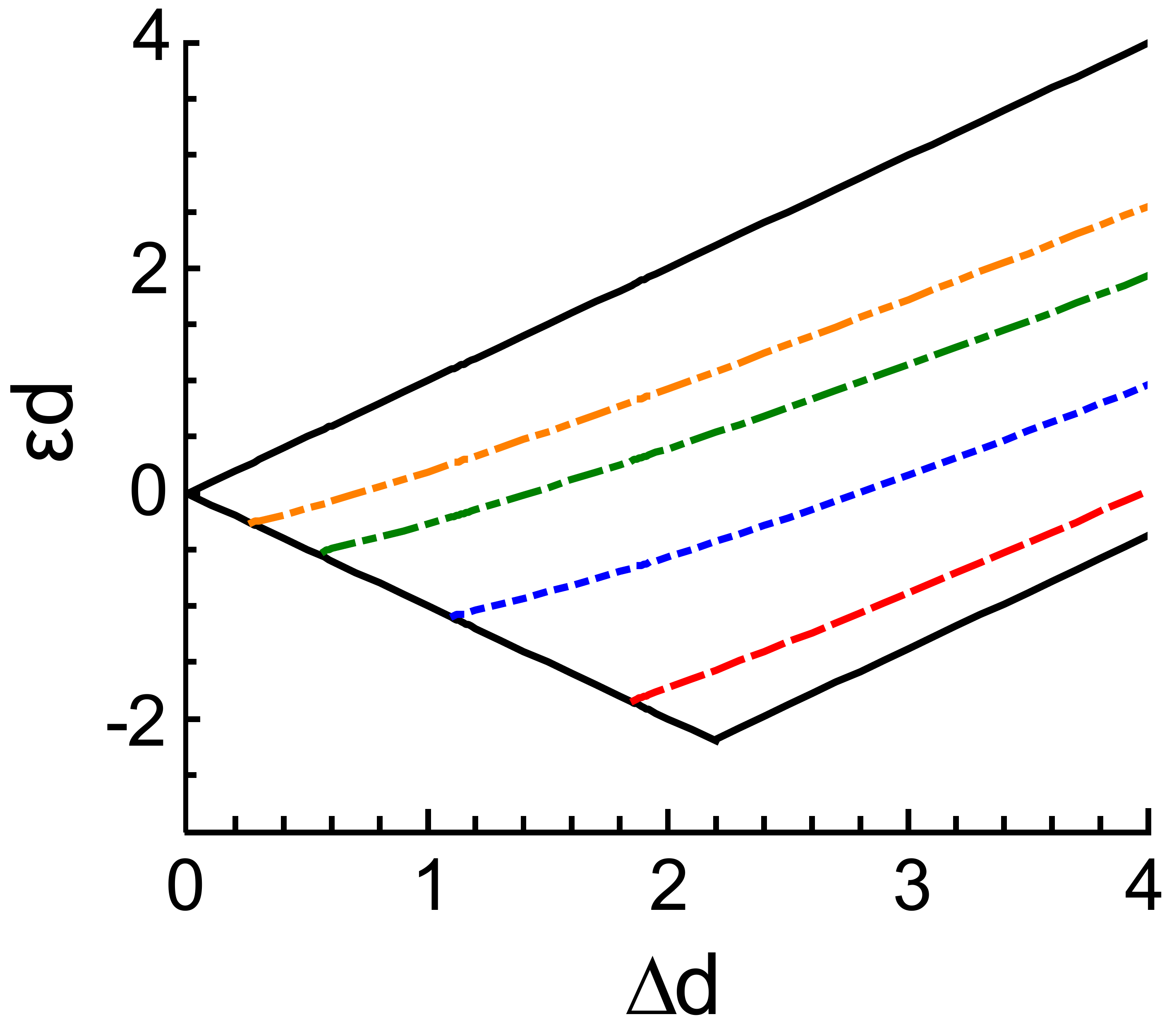}
 \caption{(Color online) A plot of the dependence of the bound state energies on band gap for the lowest four states: the ground state (dashed red line), the first excited state (dotted blue line), the second state (dot-dashed green line) and the third state (dot-dot-dashed orange line) respectively, where $U_0 = \tfrac{300}{137}$. The solid black lines denote the region bound states must fall between.}
 \label{energy}
\end{figure}

Imposing continuity on the wavefunction components at $x= \pm d/2$ leads to the following transcendental equation governing the energy quantization of bound states
\begin{equation}
\label{truncated11}
  1 - \lambda_{+}/\lambda_{-} = 0,
\end{equation}
where
\begin{equation}
\label{truncated12}
 \lambda_{\pm} = \frac{\tfrac{k}{\Delta \tau_{\pm}} \tan \left( \tfrac{kd}{2}\right) \pm \eta_{\pm}}{\eta_{\pm} \tan \left( \tfrac{kd}{2}\right) \mp  \tfrac{k}{\Delta \tau_{\pm}}},
\end{equation}
\begin{equation}
\label{truncated13}
 \eta_{\pm} = i \left( \frac{\varepsilon + 2 U_0 /d}{\Delta \tau_{\pm}} \right) \mp 1,
\end{equation}
\begin{equation}
\label{truncated14}
 \tau_{\pm} = \left( \frac{\kappa+ i \varepsilon}{\Delta} \frac{W_{\mu, \nu+1} (\kappa d)}{W_{\mu, \nu} (\kappa d)} \right)^{\pm 1},
\end{equation}
which can be solved via the usual root-searching procedures. In Fig.~\ref{stone1} we plot electron densities for the four lowest bound states for $\Delta d = 1$ and $U_0 = \tfrac{300}{137}$. Most noticeable is the absence of the single-peaked and double-peaked electron densities (the naturally expected ground and first excited states). This is because for the chosen value of the bandgap there are no such solutions to Eq.~\eqref{truncated11} inside the allowed region of bound states, as represented graphically in Fig.~\ref{energy}, where we show only the four lowest states for clarity, whereas there is an infinite number of bound states for any value of bandgap energy. The critical bandgap energies, below which the three lowest bound states are lost into the continuum, are $(\Delta d)_c = 1.86, 1.11, 0.57$. As one further decreases $\Delta d$ successively higher bound states are lost one after another. The disappearance of low-lying states from the discrete spectrum is a generic feature of the Coulomb potential independent of its regularization at small distance: in the case of Sec.~\ref{sec2}, one finds the lowest three states merge with the continuum at $(\Delta a)_c = 0.56, 0.23, 0.10$.

Lower energy bound states diving into the continuum below the bandgap is a signature of the so-called atomic collapse \cite{Schiff, Zeldovich}. Its appearance in 1D Dirac materials, with its dependence on critical bandgaps, opens a new avenue to explore such an exotic relativistic quantum mechanical phenomena in a tabletop experiment. In fact, quasi-1D Dirac systems, like carbon nanotubes, are arguably more suitable for table-top experiments on atomic collapse than graphene. Unlike graphene with a 2D Coulomb potential, the system considered here contains a band gap, which can even be controlled by external electric \cite{Li, Gunlycke} or magnetic \cite{Ajiki, Fedorov, Portnoi} fields, and admits truly bound state solutions with square-integrable wavefunctions. In gapless graphene, confinement in 2D radial trapping potentials is only possible at zero-energy \cite{Zero}.

The results shown in Fig.~\ref{energy} are somewhat similar to those found in graphene for bound states in a 1D square potential well extended infinitely in the $y$-direction, with the role of the bandgap being played by the longitudinal wavevector $\Delta \to k_y$ \cite{Pereira, Tudorovskiy}. The most important difference is that the Coulomb problem admits an infinitely large family of bound states for every nonzero size of the bandgap, albeit some deeper states may be missing for small bandgap energies. In the square well, which is in fact of less practical relevance due to the difficulty in creating sharp potential barriers in realistic graphene-based devices, as the bandgap (or the momentum $k_y$ along the quantum well) gets smaller so does the finite number of bound states present beyond the continuum. This difference in the structure of energy levels near the band edge is crucial for understanding the influence of excitonic effects on optical spectra in quasi-1D systems \cite{Banyai, Haug}.

\section{\label{conc}Conclusion}

We have presented the exact solutions to the quasi-relativistic shifted and truncated Coulomb problems for a quasi-relativistic 1D matrix Hamiltonian, which has a direct application to the growing research area of Dirac materials \cite{Wehling}.

We have shown that manipulating the size of the bandgap allows one to exclude from the discrete spectrum certain low-lying quantum states, for example the ground state, in stark contrast to the non-relativistic case. The bandgap can be controlled, e.g. in the case of carbon nanotubes, by applying an external fields \cite{Li, Gunlycke, Ajiki, Fedorov, Portnoi} or via strain \cite{Nicholas}; or in graphene nanoribbons by choosing certain nanoribbons with a desirable geometry \cite{Brey}. Alternatively, the strength of the interaction potential can be controlled by having multiple charged impurities \cite{Wang} or changing the dielectric environment.

We hope some interesting features arising from Coulomb physics, such as atomic collapse effects, can soon be observed either in the currently known quasi-1D Dirac materials or in future crystals synthesized with the latest techniques \cite{Siidra}.

\section*{Acknowledgments}
This work was supported by the UK EPSRC (CAD), the EU FP7 ITN NOTEDEV (Grant No. FP7-607521), and FP7 IRSES projects CANTOR (Grant No. FP7-612285), QOCaN (Grant No. FP7-316432), and InterNoM (Grant No. FP7-612624). We would like to thank Vit\'{o}ria Carolina for fruitful discussions and Tom Bointon for a careful reading of the manuscript. CAD appreciates the hospitality of the 08:51 First Great Western service to London Paddington where some of this work was carried out.

\end{document}